\documentclass[12pt]{article}
\usepackage{graphicx, url}
\usepackage{palatino}
\usepackage{color}
\usepackage{amsmath, amssymb,bbm}
\usepackage{epsfig}
\usepackage{rotating}
\usepackage{dcolumn}
\usepackage{bm}
\newcommand{\comment}[1]{}
\def \non{\nonumber}
\def \ra{\rightarrow}
\def \bea{\begin{eqnarray}}
\def \eea{\end{eqnarray}}

\def \ccbar{c\overline{c}}

\def \sg{\sigma}

\begin{document}   
\baselineskip 18pt
\title{{\boldmath $\chi_c$} production in modified NRQCD}
\author{
   Sudhansu~S.~Biswal$^1$\footnote{E-mail: sudhansu.biswal@gmail.com}, 
   Sushree~S.~Mishra$^1$\footnote{Email: sushreesimran.mishra97@gmail.com}
     ~and  K.~Sridhar$^2$\footnote{E-mail: sridhar.k@apu.edu.in} \\ [0.2cm]
    {\it \small 1. Department of Physics, Ravenshaw University,} \\ [-0.2cm]
    {\it \small Cuttack, 753003, India.}\\ [-0.2cm]
    {\it \small 2. School of Arts and Sciences, Azim Premji University,} \\ [-0.2cm]
    {\it \small Sarjapura, Bangalore, 562125, India.}\\
}
\date{}
\maketitle
\begin{abstract}
\noindent In a previous paper, we had modified Non-Relativistic QCD as it applies to quarkonium
production by taking into account the effect of perturbative soft-gluon emission from the colour-octet
quarkonium states. We tested the model by fitting the unknown non-perturbative parameter in the model from
Tevatron data and using that to make parameter-free predictions for $J/\psi$ and $\psi '$ production at the LHC. 
In this paper, we study $\chi_c$ production: we fit as before the unknown matrix-element using data from
Tevatron.  We, then, extend the results of the previous paper for $J/\psi$ production by calculating the effect
of $\chi_c$ feed-down to the $J/\psi$ cross-section, which, by comparing with CMS results at $\sqrt{s}=$ 13 TeV,
 we demonstrate to be small.  We have also computed $\chi_c^1$ and $\chi_c^2$ at $\sqrt{s}=$7
TeV and find excellent agreement with data from the ATLAS experiment.
\end{abstract}
\maketitle

\noindent The effective theory for studying heavy quarkonium physics,
Non-Relativistic Quantum Chromodynamics (NRQCD)~\cite{bbl}, found much
success in explaining the systematics of charmonium production at the
Fermilab Tevatron \cite{cdf} in contrast to the then existing model of quarkonium
production -- the colour-singlet model \cite{br}. But while NRQCD predicted 
\cite{polar1,polar2} a fully transversely polarised $J/\psi$ at large $p_T$ 
the Tevatron experiments found no evidence for this \cite{polar3}. These
and other problems like the $\eta_c$ cross-section measured in the LHCb
experiment \cite{lhcb} suggest that NRQCD may require some modification
to address quarkonium production fruitfully.

One approach exploits the fact that 
the colour-singlet model predicts zero polarisation and attempts to increase
the colour-singlet contribution by invoking higher-order effects
in the singlet channel \cite{gong, artoisenet}. 
\footnote{For reviews of the status of these calculations and their experimental 
consequences, see Refs.~\cite{lansberg1, lansberg2}.} For this approach to work
one has to show why colour-octet operators are small in the regions of phase space
under consideration.

In NRQCD, colour-singlet or colour-octet $c \bar c$ states are produced at the
perturbative level and these states then transform into the physical charmonium
state by means of a non-perturbative transition. The separation of the short-distance
perturbative process from the non-perturbative part is given by the factorization
theorems proved within NRQCD. The dynamics of NRQCD also helps us keep track
of the quantum numbers of the colour-octet (or singlet) $c \bar c$ state that
transforms into the physical charmonium state. 

The cross section for production of a quarkonium state $H$ is given as
\bea
  \sigma(H)\;=\;\sum_{n=\{\alpha,S,L,J\}} {F_n\over {M_Q}^{d_n-4}}
       \langle{\cal O}^H_n({}^{2S+1}L_J)\rangle, 
\label{factorizn}
\eea
where $F_n$'s are the short-distance coefficients and ${\cal O}_n$ are 
operators of naive dimension $d_n$, describing the long-distance effects.  
These non-perturbative matrix elements are guaranteed to be
energy-independent due to the NRQCD factorization formula, so that they
may be extracted at a given energy and used to predict quarkonium cross-sections
at other energies. Other than prediction of the energy-dependence of the $J/\psi$
$p_T$ distribution, several independent tests of the effective theory have been proposed \cite{tests}.

In a recently proposed modification of NRQCD \cite{bms}, we had suggested that
the colour-octet $c \bar c$ state can radiate several 
soft {\it perturbative} gluons -- each emission taking away little energy but carrying 
away units of angular momentum. In the multiple emissions that the colour-octet state
can make before it makes the final NRQCD transition to a quarkonium state, the
angular momentum and spin assignments of the $c \bar c$ state changes constantly.

For $J/\psi$, for example, the NRQCD cross-section formula which was given as follows when
written down explicitly in terms of the octet and singlet states
\begin{eqnarray}
\sigma_{J/\psi}  = \hat F_{{}^{3}S_1^{[1]}} \times \langle {\cal O} ({}^{3}S_1)^{[1]}) \rangle +
                \hat F_{{}^{3}S_1^{[8]}} \times \langle {\cal O} ({}^{3}S_1)^{[8]}) \rangle +\cr
                 \hat F_{{}^{1}S_0^{[8]}} \times \langle {\cal O} ({}^{1}S_0)^{[8]}) \rangle 
                + {1 \over M^2} \biggl\lbrack\hat F_{{}^{3}P_J^{[8]}} \times \langle {\cal O} 
                     ({}^{3}P_J)^{[8]} \rangle \biggr\rbrack .
\label{Fock}
\end{eqnarray}
gets modified to the following in the modified NRQCD with perturbative soft gluon emission:
\begin{eqnarray}
\sigma_{J/\psi} &=& \biggl\lbrack \hat F_{{}^{3}S_1^{[1]}} 
                \times \langle {\cal O} ({}^{3}S_1)^{[1]}) \rangle \biggr\rbrack \cr 
                &+& \biggl\lbrack  
                  \hat F_{{}^{3}S_1^{[8]}} 
                 + \hat F_{{}^{1}P_1^{[8]}} 
                + \hat F_{{}^{1}S_0^{[8]}} + (\hat F_{{}^{3}P_J^{[8]}} ) \biggr\rbrack 
                \times ({\langle {\cal O} ({}^{3}S_1)^{[1]} \over 8}) \rangle \cr
                &+& \biggl\lbrack  
                  \hat F_{{}^{3}S_1^{[8]}} 
                 + \hat F_{{}^{1}P_1^{[8]}} 
                + \hat F_{{}^{1}S_0^{[8]}} + (\hat F_{{}^{3}P_J^{[8]}} ) \biggr\rbrack 
                \times \langle {\cal O}  \rangle ,
\label{modified1}
\end{eqnarray}
where
\begin{equation}
     \langle {\cal O}  \rangle =
                \times \biggl\lbrack 
                 \langle {\cal O} ({}^{3}S_1^{[8]}) \rangle 
                + \langle {\cal O} ({}^{1}S_0^{[8]}) \rangle 
                + {\langle {\cal O} ({}^{3}P_J^{[8]}) \rangle \over M^2}
                    \biggr\rbrack.
\end{equation}

In contrast to the usual case, where we needed to fix three non-perturbative
parameters to get the $J/\psi$ cross-section, in our case it is the sum of these
parameters: so we have a single parameter to fit. 

In Ref.~\cite{bms}, $J/\psi$ and $\psi '$ production in modified NRQCD was already
studied. For both these charmonium states, the non-perturbative parameter (the
single one that we needed to fit) was fitted from the old Tevatron data \cite{cdf} and
we used the fitted parameter to make {\it predictions} for $J/\psi$ and $\psi '$
production at the LHC and our predictions were in excellent agreement with the
data. For $J/\psi$ production at the LHC, the CMS experiment \cite{cms} does
not distinguish between direct $J/\psi$ and those coming from $\chi_c$ states. We
had in that case only made a rough estimate of the magnitude of the $\chi_c$
contribution and were convinced that it was small. So the theoretical results we compared with
CMS data in Ref.~\cite{bms} were the direct production ones.

To do the full $J/\psi$ production at LHC we need to also include the result of the
feed-down from the $\chi_c$ states and we undertake this task in this paper. To do
this we need to extract the non-perturbative parameter for $\chi_c$ production from
a fit to Tevatron data and then use that to compute the $\chi_c$ production at LHC
and its contribution to the $J/\psi$ cross-section. The ATLAS experiment has also
measured and presented results on $\chi_c^1$ and  $\chi_c^2$ production \cite{ATLAS} at 7 TeV
 energy. We are in a position to also compare our results with these data.

For $\chi_c$ production the cross-section expression, similar to Eq.~\ref{modified1} is given
by:

\begin{eqnarray}
\sigma_{\chi_c} &=& \biggl\lbrack \hat F_{{}^{3}P_J^{[1]}} 
                \times \langle {\cal O}^{\chi} ({}^{3}P_J^{[1]}) \rangle \biggr\rbrack \cr 
                &+& \biggl\lbrack  
                  \hat F_{{}^{3}S_1^{[8]}} 
                 + \hat F_{{}^{1}P_1^{[8]}} 
                + \hat F_{{}^{1}S_0^{[8]}} + \hat F_{{}^{3}P_J^{[8]}}  \biggr\rbrack 
                \times ({{\langle {\cal O}^{\chi} ({}^{3}P_J^{[1]}) } \rangle \over 8} ) \cr
                &+& \biggl\lbrack  
                  \hat F_{{}^{3}S_1^{[8]}} 
                 + \hat F_{{}^{1}P_1^{[8]}} 
                + \hat F_{{}^{1}S_0^{[8]}} + \hat F_{{}^{3}P_J^{[8]}}  \biggr\rbrack 
                \times \langle {\cal O}^{\chi}  \rangle,  
\label{modified2}
\end{eqnarray}
where
\begin{equation}
     \langle {\cal O}^{\chi}  \rangle =
                \times \biggl\lbrack { M^2
                 \langle {\cal O} ({}^{3}S_1^{[8]}) \rangle} 
                + {M^2 \langle {\cal O} ({}^{1}S_0^{[8]}) \rangle} 
                + {\langle {\cal O} ({}^{3}P_J^{[8]}) \rangle}
                    \biggr\rbrack. 
\end{equation}

The cross-section kinematics are familiar:
\bea
&&\frac{d\sg}{dp_{_T}} \;(p \bar p \ra \ccbar\; [^{2S+1}L^{[1,8]}_J]\, X)= \non \\
&&\sum \int \!dy \int \! dx_1 ~x_1\:G_{a/p} (x_1)~x_2\:G_{b/p}(x_2) 
\:\frac{4p_{_T}}{2x_1-\overline{x}_T\:e^y}\non\\
&&\frac{d\hat{\sg}}{d\hat{t}}
(ab\ra \ccbar[^{2S+1}L_J^{[1,8]}]\;d),
\label{eq:diff}
\eea
where the summation is over the partons ($a$ and $b$),    
$G_{a/p}$,   
$G_{b/p}$ are the distributions of partons $a$ and $b$ in the
protons and $x_1$, $x_2$ are the respective  momentum they carry.
In the above formula, $\overline{x}_T=\sqrt{x_T^2+4\tau} \equiv 2 M_T/\sqrt{s}$ \ with \  
$x_T=2p_{_T}/\sqrt{s}$ and  \(\tau=M^2/s\).
$\sqrt{s}$ is the center-of-mass energy, 
$M$ is the mass of the resonance and $y$ is the rapidity at which the 
resonance is produced. 
The matrix  elements for the subprocesses can be found in Refs.~\cite{cho2,gtw, Mathews}. 

We use, as we did with $J/\psi$ and $\psi '$ in Ref.~\cite{bms}, the Tevatron data
to determine the non-perturbative parameter for $\chi_c$ production. However, the
CDF experiment at the Tevatron does not give us the individual cross-sections for
the three states $\chi_c^{0,\ 1,\ 2}$ but only the sum of all three resonances all
decaying into the $J/\psi$. We assume that the non-perturbative parameters for the
three states are equal and with this assumption we need to extract only a single
parameter from the $\chi_c$ $p_T$ distribution that the CDF experiment has made
available \cite{cdf}. The fit to the CDF $\chi_c$ distribution is shown in Fig. 1.

\begin{figure}[h!]
\begin{center}
\includegraphics[width=8cm,height=8cm]{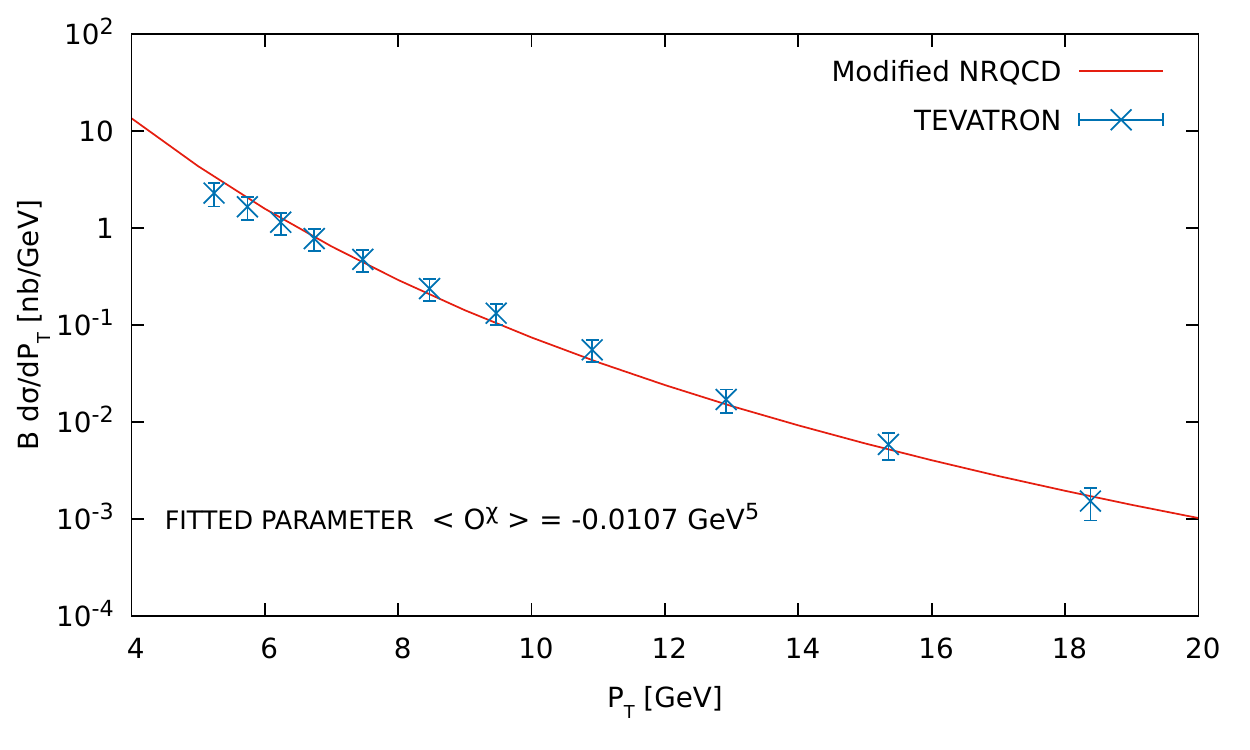}
\caption{Predicted differential cross section 
fitted to the data on $\chi_c$ production from the CDF experiment
at Tevatron at $\sqrt{s} = 1.8$ TeV. }
\label{fig:1}
\end{center}
\end{figure}

With the non-perturbative parameter so obtained from the CDF experiment, we are
now in a position to calculate the $p_T$ distributions for each of the $\chi_c$ states
at LHC energies and, after folding in the branching ratios for these states to
decay into a $J/\psi$, we are able to calculate the inclusive $J/\psi$ $p_T$ distribution
with the $\chi_c$ contribution included and compare with the 13 TeV data from the
CMS experiment (see Fig.~2).  The experiment has made measurements in five different rapidity
intervals and we have also carried out our computation for all five intervals. In Fig.~2, where
we have shown the comparison of our results with those of the experiment, we have the theoretical
results for inclusive as well as direct $J/\psi$ production. True to the estimates we had made earlier, 
the contribution to the $J/\psi$ cross-section from $\chi_c$ feed-down is very small.

\begin{figure}[h!]
\begin{center}
\includegraphics[width=8cm,height=8cm]{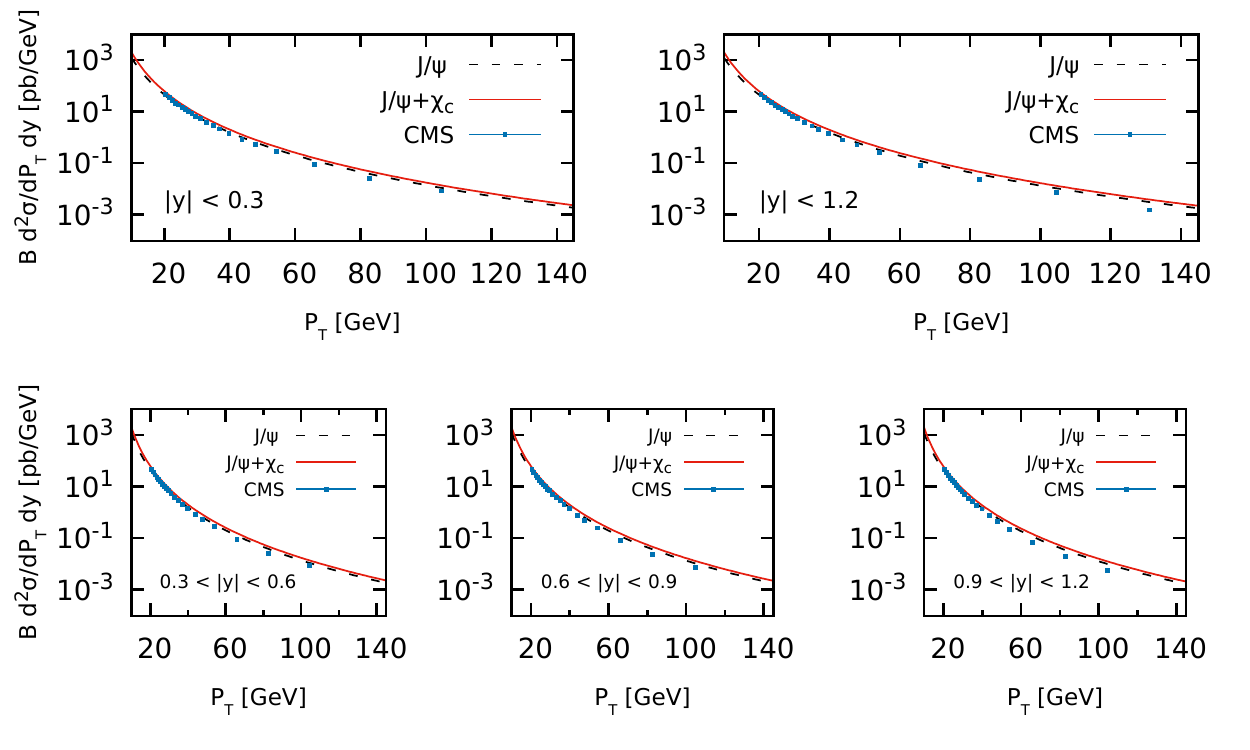}
\caption{Predicted differential distributions for
full $J/\psi$ production at the LHC running at
13 TeV compared  with data
from the CMS experiment.}
\label{fig:2}
\end{center}
\end{figure}

\begin{figure}[h!]
\begin{center}
\includegraphics[width=8cm]{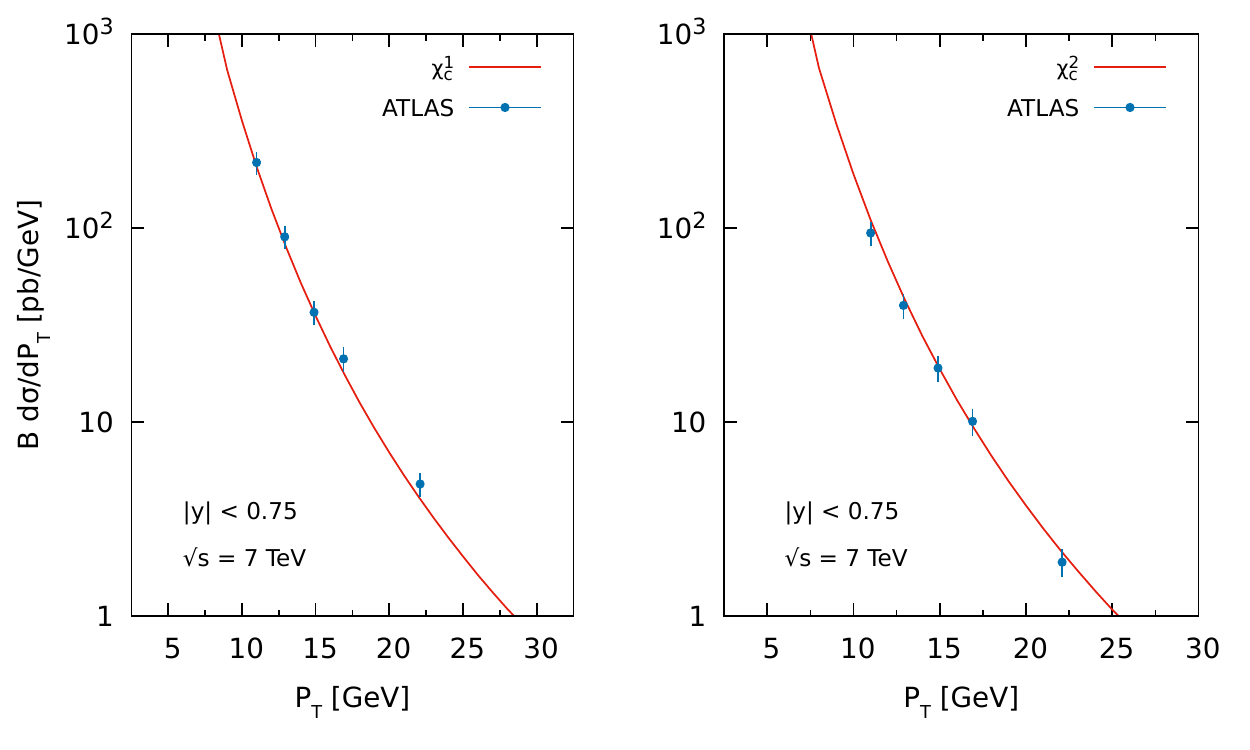}
\caption{ Predicted differential distributions for
$\chi_c^1$ and $\chi_c^2$ production at the LHC running at
 7 TeV compared  with data
from the ATLAS experiment.}
\label{fig:3}
\end{center}
\end{figure}
The ATLAS experiment at the LHC has data on $\chi_c^1$ and $\chi_c^2$ production at
at a $\sqrt{s}$ of 7 TeV. A comparison of our theoretical predictions with these states
provides the most direct check on our theoretical model. In Fig.~3, we show the comparison
of our predictions with the ATLAS data for both $\chi_c^1$ and $\chi_c^2$ and a very good
agreement between our model predictions and the data is seen to result. 


To conclude, following up on the success of the modified NRQCD that we proposed in Ref.~\cite{bms}
in predicting the data on $J/\psi$ and $\psi '$ at the LHC, we calculated the cross-section for
$\chi_c$ production in this paper. Using the Tevatron data to fit the single non-perturbative parameter
that we need, we then used it to compute inclusive $J/\psi$ production at $\sqrt{s}$=13 TeV and compared
it with the data from the CMS experiment. Our results show that the contribution to the $J/\psi$ cross-section
from $\chi_c$ feed-down is not significant. We have also computed $\chi_c^1$ and $\chi_c^2$ a at $\sqrt{s}=$7
TeV and a comparison with the data from the ATLAS experiment are seen to be very good.

%

\section*{Acknowledgments}
We would like to thank Vaia Papadimitriou for comments and discussions.


\end{document}